# Daniel Mögling's sunspot observations in 1626 – 1629: A manuscript reference for the solar activity before the Maunder Minimum

Hisashi Hayakawa (1-4)*, Tomoya Iju (5), Koji Murata (2, 6), Bruno P. Besser (7-8)

(1) Institute for Space-Earth Environmental Research, Nagoya University, Nagoya, 4648601, Japan

(2) Institute for Advanced Research, Nagoya University, Nagoya, 4648601, Japan

(3) UK Solar System Data Centre, Space Physics and Operations Division, RAL Space, Science and Technology Facilities Council, Rutherford Appleton Laboratory, Harwell Oxford, Didcot, Oxfordshire, OX11 0QX, UK

(4) Nishina Centre, Riken, Wako, 3510198, Japan

(5) National Astronomical Observatory of Japan, 1818588, Mitaka, Japan

(6) Graduate School for Humanities, Nagoya University, Nagoya, 4648601, Japan

(7) Space Research Institute, Austrian Academy of Sciences, Graz, 8042, Austria

(8) Institute of Physics, University of Graz, Universitätsplatz 5/II, 8010 Graz, Austria

* hisashi@nagoya-u.jp

**Abstract**

The sunspot groups have been observed since 1610 and their numbers have been used for evaluating the amplitude of solar activity. Daniel Mögling recorded his sunspot observations for more than 100 days in 1626 – 1629 and formed a significant dataset of sunspot records before the Maunder Minimum. Here, we have analysed his original manuscripts in the Universitäts- und Landesbibliothek Darmstadt (ULBD) to review Mögling's personal profile and observational instruments and derive number and positions of the sunspot groups. In his manuscript, we have identified 134 days with an exact sunspot group number and 3 days of additional descriptions. Our analyses have completely revised their observational dates and group number, added 19 days of hitherto overlooked observations, and removed 8 days of misinterpreted observations. We have also revisited sunspot observations of Schickard and Hortensius and revised their data. These results have been compared with the contemporary observations. Moreover, we have derived the sunspot positions from his sunspot drawings and located them at 2°–23° in the heliographic latitude in both solar hemispheres. Contextualised with contemporary observations, these results indicate their temporal migration to lower heliographic latitudes and emphasise its location in the declining phase of Solar Cycle −12 in the 1620s. His observations were probably conducted using a pinhole and camera obscura, which made Mögling likely underestimate the sunspot group number by ≥ 33% – 52 %. This underestimation should be noted upon their comparison with the modern datasets.

1. Introduction

Daily records of sunspot observations, which have formed an essential basis for evaluating long-term solar activity since 1610, have often been considered as one of the longest ongoing scientific experiments in modern science (Owens, 2013; Vaquero *et al*., 2016; Arlt and Vaquero, 2020). After the initial modern compilation of their comprehensive dataset in Hoyt and Schatten (1998a, 1998b = HS98), recent studies have continuously recalibrated and improved these data series to revise the overall long-term trend (*e.g.*, Clette *et al*., 2014; Clette and Lefèvre, 2018). Investigations of the original observational records have formed the basis for these analyses (Vaquero *et al*., 2011, 2016; Arlt *et al*., 2013; Usoskin *et al*., 2015; Carrasco *et al*., 2016; Svalgaard, 2017). They have offered a ground truth for further recalibrations using sophisticated methods (Vaquero *et al*., 2016 = V+16; Clette and Lefèvre, 2018). However, as depicted in Figure 2 of Muñoz-Jaramillo and Vaquero (2019), it is challenging to extend these analyses beyond the mid-nineteenth century and their reconstructions remain controversial (Clette and Lefèvre, 2018; Svalgaard and Schatten, 2016; Usoskin *et al*., 2016, 2021; Chatzistergos *et al*., 2017; Willamo *et al*., 2017), especially toward and beyond the Maunder Minimum (*e.g.*, Usoskin *et al*., 2015; Vaquero *et al*., 2015; Zolotova and Ponyavin, 2015).

Even after the compilation of the revised database for historical sunspot observations (V+16), such reanalyses are ongoing efforts that have modified a number of historical observational datasets (*e.g*., Arlt, 2018; Hayakawa *et al*., 2018a, 2018b; Carrasco *et al*., 2019a, 2019b; Karoff *et al*., 2019), including long-term observations around the Maunder Minimum (*e.g*., Carrasco *et al*., 2019c; Hayakawa *et al*., 2020a, 2021) and the Dalton Minimum (Hayakawa *et al*., 2020b). Sunspot drawings of the early 17th century are of particular interest, as they provide unique evidence for the solar activity before the Maunder Minimum, and moderate revisions or additions can update the existing understanding (*e.g.*, Vaquero *et al*., 2011; Carrasco *et al*., 2019a). To date, the major observers' observational records before the Maunder Minimum have been recently analysed to improve sunspot group number and derive sunspot positions (Arlt *et al*., 2016; Vokhmyanin and Zolotova, 2018a, 2018b; Carrasco *et al*., 2019a, 2019b, 2020; Vokhmyanin *et al*., 2020, 2021). These results have characterised solar cycles before the Maunder Minimum, both with the sunspot group number and butterfly diagrams, clarified their significant discontinuity with the Maunder Minimum, and formed a basis for improving solar dynamo models (*e.g.*, Hotta *et al.*, 2019; Charbonneau, 2020). On the other hand, their sparse availability requires further data to improve their reconstructions (Muñoz-Jaramillo and Vaquero, 2019).

In this context, little is known about Daniel Mögling's sunspot observations, whereas this observer, labelled as "Mogling" in HS98 and V+16, has been considered the 5th most active sunspot observer before the onset of the Maunder Minimum, following Scheiner, Hevelius, Harriot, and Malapert (see HS98 and V+16) and even more active than Galilei and other contemporary observers (*e.g.*, Vokhmyanin and Zolotova, 2018a, Carrasco *et al.*, 2020). His observations span 1626 – 1629 and form one of important datasets during Solar Cycle −12 spanning in the 1620s (*e.g.*, Figure 27 of Arlt and Vaquero, 2020). Locating his autograph manuscript at the Universitäts- und Landesbibliothek Darmstadt (ULBD), we analysed his sunspot observations and clarified his data and metadata. Here, we first profiled his biographical background, observational instruments, and philological details on his observational records (Section 2). We then analysed his observational records to derive sunspot group numbers in the Waldmeier classification to revise the existing data and include forgotten data (Section 3). Using the revised data, we also derived the sunspot positions recorded in his sunspot drawings (Section 4). We have summarised and contextualised these results in comparison with contemporary sunspot observations, including those of Schickard and Hortensius, revised in this article (Section 5).

## 2. Daniel Mögling and his Observations

Daniel Mögling (1596 – 1635) was born in Böblingen near Stuttgart and raised by his mother because his father, a physician, passed away soon after his birth in an epidemic. Members of his family had been professors at the University of Tübingen for several generations. Daniel entered the same university in April 1611, where he received his bachelor's and master's degrees in philosophy in September 1612 and February 1615, respectively. After a year of academic peregrination, he enrolled in medical studies at the University of Altdorf near Nürnberg. He then returned to the University of Tübingen at the end of 1618, and finished his studies in 1621. His interests also extended to physics and, in particular, astronomy. In May 1621, he started his almost lifelong occupation as a court physician under Philipp III (1581 – 1643), Landgrave of Hessen-Butzbach. According to his employment contract, Mögling was also required to work on mathematics and astronomical observations (Neumann, 1995). During the course of his profession, he got in touch with several important contemporary astronomers, such as Wilhelm Schickard (1592 – 1635) and Johannes Kepler (1571 – 1630). In fact, Kepler visited his observatories at Butzbach (N50°26, E8°40; see also Figure 1) at least twice, in July 1621 and September 1627 (Rösch, 1975). The contract between Mögling and Landgrave was cancelled in 1635, probably because of the approach of the Thirty Years War to Butzbach. Landgrave Philipp recommended Mögling to his nephew Georg II of Hessen-Darmstadt for a position at the University of Marburg (today Marburg an der Lahn), but before its realisation, Mögling died of pestilence in August 1635 in Butzbach.

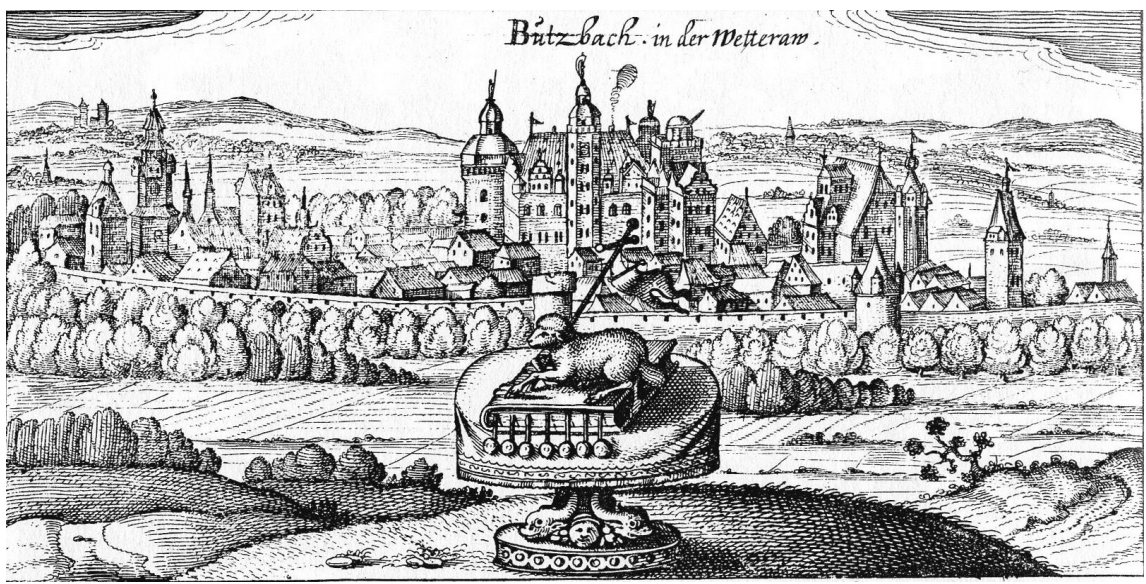

Figure 1: Merian's copper print of one of Mögling's observatories at Butzbach, adopted from Zeiller and Merian (1646). His observatory was located on the right side of the castle, with the balustrade around the tower with spherical top. Besides, he had placed a large telescope in the other observatory in the garden (Rösch, 1975; Rößling, 2010; Popplow, 2013).

Mögling had at least three instruments for solar observations, according to his inventory manuscript dated November 1628 (ULBD HS[1] 3020). Here, we have located descriptions on a black wooden tube for observing the solar radii (No. 87 in ULBD HS 3020, *f*.[2] 15a), gilded sphere with an eye-tube for solar spots (No. 123 in ULBD HS 3020, *f.* 15b), and silver-style rod eye-tube, 4 feet long, for observing the Sun's position (No. 151 in ULBD HS 3020, *f.* 16a). Accordingly, we consider that Mögling measured the Sun's position and radii with instruments No. 151 and 87, and monitored sunspots with instrument No. 123.

---

[1] HS is an abbreviation of "*Handschrift* (manuscript)" used as a part of shelf mark.
[2] Here we describe a singular folio as "*f.*".

Kepler's description allows us to confirm this supposition and even indicates Mögling's records of sunspots shown in the projected images. Upon his visit to Butzbach in 1627, Kepler stated, "In an open and spacious place, a thirty-foot-high stake is fixed; at the top, a pulley is placed, through which a capstan cable is passed and it surrounds a fifty-foot-long tube, driven with great difficulty by six robust men from its ridge; this tube is raised to such a height that, through its hole, which is the size of a pea, or a lens, or even a grain of millet, the Sun projects its rays onto an opposite white shelf, which terminates the cavity of the tube at its bottom. On the tablet, then, one can clearly distinguish the sunspots, which are formed by the simple hole, without the interposition of any convex glass" (Kepler, 1629, p. 4; Kepler, 1983, p. 469; see also Jeandillou and Mehl, 2018, p. 68). Kepler further describes Mögling's interest in the motions of sunspots, their seasonal inclinations, and their existence on the solar surface. Kepler witnessed his sunspot drawings showing their motions.

Mögling's drawings were later compiled in a manuscript "*Observationes macularum Solis* (Observations of sunspots)", which is currently preserved in the ULBD as HS 228, probably between 1629 and 1635 by himself. This manuscript consists of 38 *folia* (hereafter, *ff.*) and involves sunspot drawings dated from 23 June 1626 to 16 June 1629. Mögling depicted two kinds of sunspot drawings: individual drawings for his daily observations and summarised drawings for motion tracking of specific sunspot groups (Figure 2), probably owing to his interest in the solar rotation period (*e.g.*, Hoyt and Schatten, 1997, p. 20). Initially, he used small sunspot drawings ($\phi^3 \approx 3.4$ cm) for his early daily observations (ULBD HS 228, *ff.* 1a – 1b) and medium sunspot drawings ($\phi \approx 7.0$ cm) for tracking the motion of specific sunspot groups (ULBD HS 228, *f.* 2a). After August 1626, he regularly depicted a large sunspot drawing ($\phi \approx 12.6$ cm) on each folio until the end of his observations.

---

[3] Here, we abbreviate "diameter" with $\phi$.

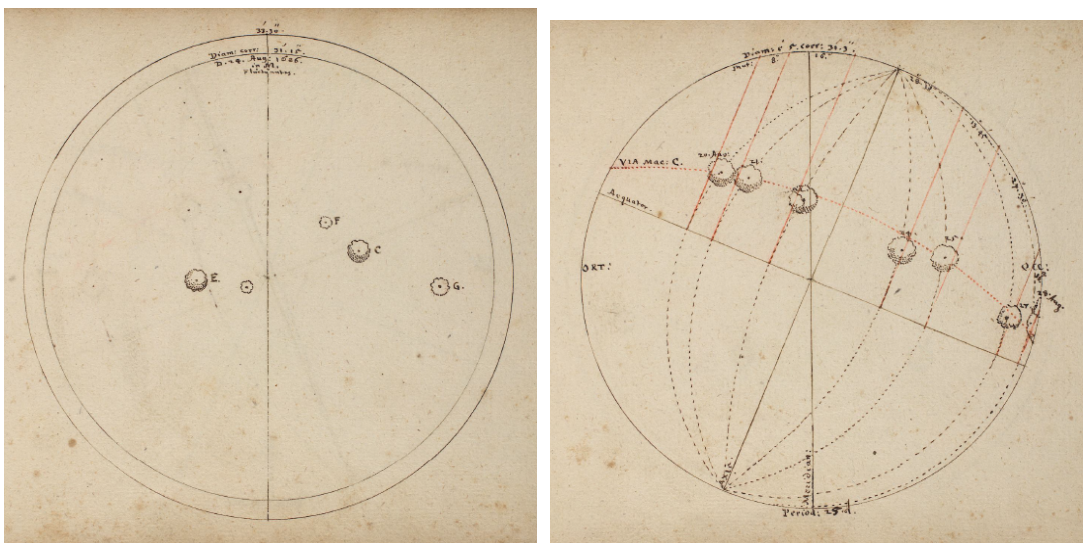

Figure 2a (left), 2b (right): Examples of Mögling's sunspot drawings for his daily observations and motion tracking of specific sunspot groups: (a) his whole-disk sunspot drawing on 3 September 1626 (ULBD HS 228, f. 8a); and (b) his motion tracking of sunspot group "C" on 30 August – 7 September 1626 (ULBD HS 228, f. 10a); © ULBD. It is to be noted that their dates, shown according to the Julian calendar, have been converted to the Gregorian calendar in this article.

### 3. Sunspot Group Number

Consulting ULBD HS 228, we acquired 137 days of Daniel Mögling's sunspot observations, applied the Waldmeier classification, and summarised their results in Figure 3[4]. His sunspot observations are found not only in his sunspot drawings (covering 103 days) but also in his textual descriptions for 34 days. Although they mostly describe spotless days, he reported several spots in three of these descriptions (16 and 20 October 1626; and 3 March 1627). As these descriptions are only small text pieces like 'several sunspots (*cum maculis* or *aliquibus maculis*)', we were able to use them just for calculations of active day fractions (ADFs), and hindered us from deriving their sunspot group number. Thus, our summary includes 134 of Mögling's datable sunspot observations, excluding these active-day reports.

Comparing our results with the existing databases (HS98 and V+16), we have realised that Mögling's observations in these datasets had been interpreted according to the Gregorian calendar and incorporated as described in his manuscript, whereas Darmstadt was under Lutheran confession and used the Julian calendar at the time of Mögling's observations (*e.g*., Gingerich, 1983). In fact, his correspondence with Schickard followed the Julian calendar, dating 23 July 1626 as Sunday, 27 September 1626 as Wednesday, and 26 February 1629 as Thursday, for example (Seck, 2002). Therefore, the dates of these observations should be converted from the Julian to the Gregorian calendar for scientific comparison with other contemporary sunspot observations. Apart from this overall calendar issue, we have included nineteen days of hitherto overlooked observations, revised three of the Julian dates, and removed 8 days of misinterpreted observations.

---

[4] https://www.kwasan.kyoto-u.ac.jp/~hayakawa/data/moegling/moegling_group_number.txt

Mögling's sunspot group number has been revised throughout and shows values lower than the existing databases (HS98 and V+16), where the individual sunspots in the same group had been occasionally split. Still, Mögling recorded multiple sunspot groups up to five in his observations and was possibly free from the arbitrary selection of the observed sunspots, which was often the case with contemporary sunspot observers (*c.f*., Carrasco *et al*., 2019a). Mögling actively recorded spotless days in 1626 – 1627 and may further benefit future analyses in ADFs in combination with other contemporary observations.

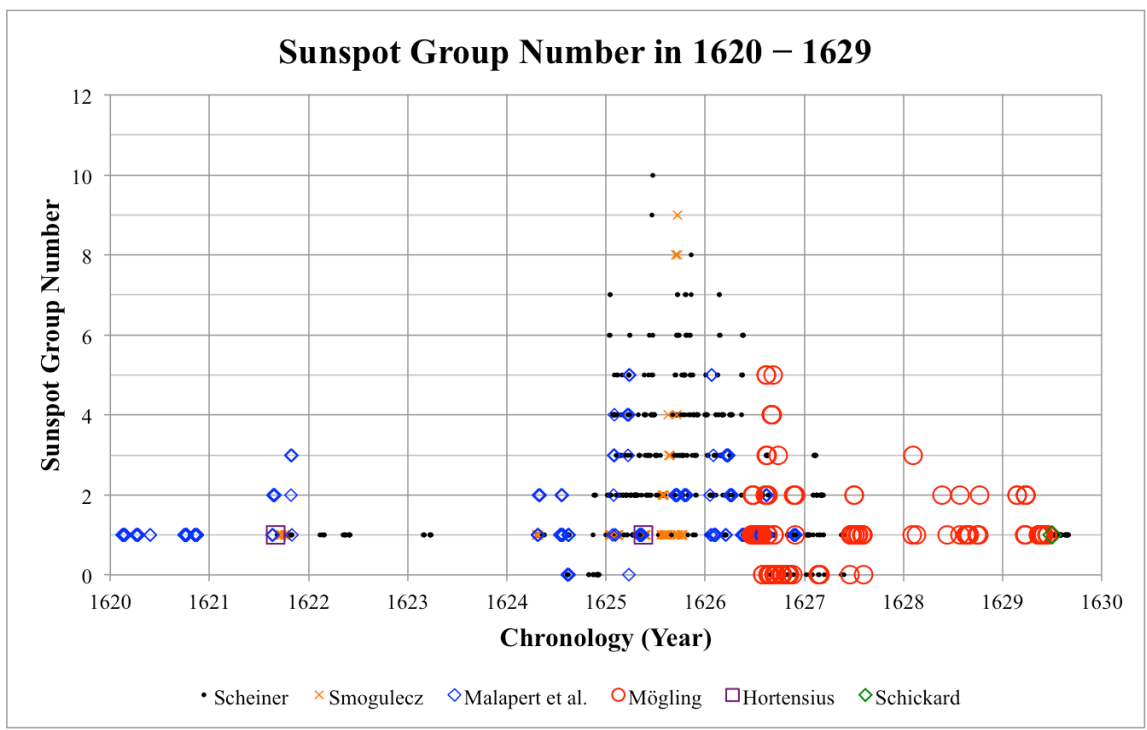

Figure 3: Sunspot group number in 1620 – 1629, based on contemporary sunspot observations. Among them, those of Mögling, Hortensius, and Schickard are revised in our study, whereas those of Malapert et al. are derived from Carrasco *et al*. (2019b), and those of Scheiner and Smogulecz are derived from Vaquero *et al*. (2016).

Mögling's data compare well with contemporary observations (*e.g.*, Vaquero *et al*., 2016; Carrasco *et al*., 2019b), as shown in Figure 3. Here, we have also consulted the records of Hortensius and Schickard, who were based at the Protestant cities of Leiden and Tübingen that used the Julian calendar at that time, to revise their observational data. This is confirmed from the dating in Schickard's correspondence, such as the identification of 29 July 1626 as Saturday and 1 October 1626 as Sunday. In the case of Hortensius, we have revised his observation to 1625 May 25 and added another observation on 20 September 1621 (Hortensius, 1633, p. 65). For Schickard, we have revised one observation to 16 July 1629 (Schickard, 1632, p. 10) and removed his observations on 9 – 11 January 1621, as they were not recorded in either of HS98's alleged sources: Schickard (1632) or Wolf (1850, p. 119). Figure 3 depicts Mögling's sunspot observations in the declining phase of Solar Cycle −12, in comparison with other contemporary observations including Schickard and Hortensius, which have been revised in this article.

**4. Sunspot Positions**

On the basis of ULBD HS 228, we also analysed the sunspot positions in Mögling's sunspot drawings to construct a butterfly diagram. As shown in Figure 2, Mögling recorded not only the daily whole-disk drawings but also drawings for the motions of specific groups of sunspots. We have combined them to derive the sunspot positions and consider that these sunspot drawings are upside-down. Despite his controversial annotations of the disk orientations around some of his sunspot drawings, the recorded sunspot groups generally move from the left to the right (see Figure 2b). This indicates the E–W orientations in his drawings set as E in the left and W in the right. Their N–S directions are inferred from the comparisons of the recorded inclinations of the sunspot motions and the $B_0$ angles for the observational dates. For instance, in early September 1626 (Figure 2b), the $B_0$ angles are calculated as ≈ 7°. This matches best with the recorded inclination of the sunspot motion when this drawing is shown upside-down. This trend has been universally confirmed throughout Mögling's sunspot drawings, as long as their sunspot motions can be tracked in chronological sequence. Therefore, it is considered that Mögling depicted sunspot drawings upside-down, with the E-W directions shown from left to right. This interpretation is consistent with Kepler's description of Mögling's instrument that projected sunspot images on a sheet of paper.

We derived the sunspot positions on this basis. We have used the scanned images of HS 228. We fitted depicted disk limbs to the circle, adjusting their disk centers. When they are geometrically distorted and either vertically or horizontally too large to be a circle, we have modified the larger diameter to fit the limbs to the circle, following the procedure of Fujiyama *et al*. (2019). We have derived the sunspot positions for the drawings that depict sunspot motion tracking of specific groups (*e.g*. Figure 2b), minimising the latitudinal deviations of each sunspot group. After deriving the sunspot positions for specific groups, we applied their estimated positions to the whole-disk drawings for his daily observations to constrain the disk orientations (*e.g*. Figure 2a). This enabled us to derive the minor sunspot groups whose motions were not tracked by Mögling himself in his manuscript ULBD HS 228.

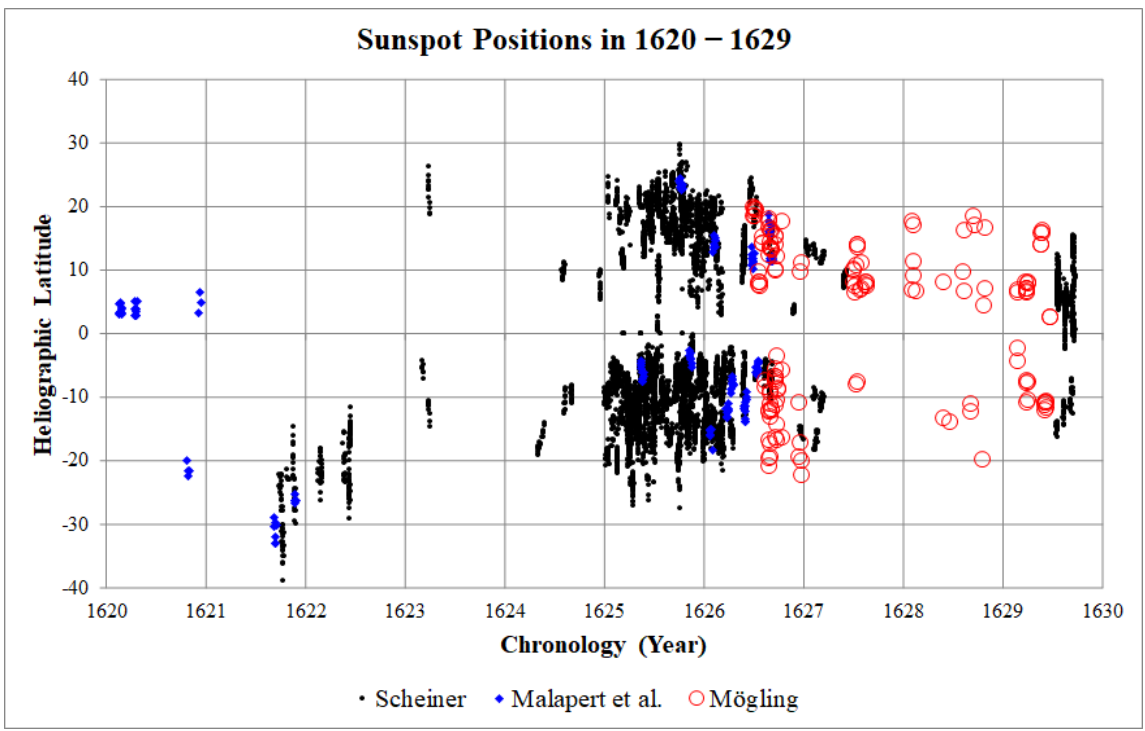

Figure 4: Sunspot positions in 1620 – 1629, consisting of the sunspot observations of Scheiner (Arlt *et al*., 2016), Malapert *et al*. (Carrasco *et al*., 2019b), and Mögling (this study).

Our results are summarised in Figure 4, which show a comparison with the existing sunspot positions derived from Scheiner and Malapert's observational accounts (Arlt *et al*., 2016; Carrasco *et al*., 2019b). Mögling's sunspot positions fill the chronological gaps in these existing datasets. In 1626, they are located at 2°–23° in both solar hemispheres. Afterward, their distributions shift more equatorward, still showing sunspots in both solar hemispheres. As such, especially in 1627 – 1629, Mögling's sunspot distributions were biased slightly more in the northern solar hemisphere and seemed to show a trend similar to Scheiner's sunspot positions in 1629 (Arlt *et al*., 2016). Their overall distributions are consistently contextualised in the declining phase of Solar Cycle −12, as confirmed by Scheiner and Malapert's accounts (Figure 4). This result contrasts this cycle with the Maunder Minimum, where most of the sunspot positions were reported only in the southern solar hemisphere (Ribes and Nesme-Ribes, 1993), as seen in the cases of Gassendi and Hevelius' sunspot positions (Vokhmyanin and Zolotova, 2018; Carrasco *et al*., 2019c).

## 5. Discussions

Mögling's manuscripts (ULBD HSs 228 and 3020) and Kepler's contemporary records indicate that his small and large drawings were the results of different instruments: the gilded sphere with the eye-tube (No. 123 in ULDB HS 3020) and the black wooden tube (No. 87 in ULBD HS 3020). The small drawings show sunspots with various time stamps without their solar diameter. This indicates that his observational instrument was likely compact and capable of tight turns. The large drawings show sunspots and solar diameters, indicating the usage of the black wooden tube for "observation of the solar radii" This is consistent with the large tube(s) documented in Kepler's report. The size of aperture was probably ≤ 1 cm, as it was compared with "the size of a pea, or a lens, or even a grain of millet", while six example sizes of tube aperture were noted in Mögling's manuscript (ULBD HSs 228, *f.* 2b; see also Jeandillou and Mehl, 2018, p. 71). Their time stamps were mostly around noon, indicating observations on a meridian line and agreeing with the heavy tube documented in Kepler's report.

Mögling's instruments for solar observations were probably not telescopes but sighting tubes without any convex lens; one of them had an aperture of ≤ 1 cm in diameter on the sun-facing side of a long tube (50 feet ≈ 12.5 m; see von Bauernfeind, 1862, p. 13) and projected the solar disk onto a white shelf placed on the other side of the tube. As such, he appears to have observed sunspots with an instrument that functioned like the camera obscura. This supposition is consistent with the upside-down orientation of the solar disk (see *e.g*. Figure 2 of Vaquero (2007)) and the somewhat blurred depiction and uneven areas of sunspots in Mögling's sunspot drawings (Figure 2). We estimate the focal lengths of his instruments as 3.9 m and 14.4 m, using the depicted aperture sizes and diameters of depicted sunspot drawings (3.4 cm and 12.6 cm). Thus, the solar images should have been projected on sheets ≈ 1.9 m away from the end of tube for the large drawings. Our estimates are consistent with how Mögling observed the solar disk: using the gilded sphere with eye-tube for the small drawings and the black wooden tube for the large drawings. Contemporary sunspot observations through pinholes have been documented in Malapert's records, whereas Malapert himself considers telescopes more suitable for sunspot observations (Carrasco *et al.*, 2019b).

These facts lead us to consider that Mögling probably missed all the small sunspots (A-type and B-type groups in the Waldmeier classification), but likely detected the larger ones (E-type, F-type, G-type, and H-type groups) and arguably the J-type sunspot groups as well. Within the modern observations, the A-type, B-type, and J-type sunspot groups account for ≈ 19%, ≈ 14%, and ≈ 19%, respectively, of all the observed sunspot groups in the dataset of the Uccle Solar Equatorial Table (USET) from 1940 – 2014 (Carrasco *et al.*, 2015). He was also unable to resolve the umbrae from the penumbrae. Therefore, it is assumed that Mögling underestimated the sunspot group number, overlooking ≥ 33% – 52 % of the total sunspot groups on the solar disk. Despite this probable underestimation, Mögling's sunspot records seem to depict the decay of the sunspot group number (Figure 3) and the migration of their positions to lower heliographic latitudes (Figure 4) from 1626 to 1629. This is typical of the declining phases of the modern solar cycles (Hathaway, 2015; Muñoz-Jaramillo and Vaquero, 2019) and agrees with the existing data for the other contemporary observations (Figures 3 and 4).

## 6. Conclusion

In this article, we analyse the sunspot observations of Daniel Mögling. His autograph has been identified with HS 228 at the ULBD. We have extended and detailed information on his life (1596 – 1635) and personal profile. He worked as a court physician/astronomer under Landgrave Philipp III of Hessen-Butzbach and recorded his astronomical observations at Butzbach (N50°26, E008°40) using at least three instruments for his solar observations. Mögling's autograph manuscript (ULBD HS 228) contains sunspot drawings from 1626 June 23 to 1629 June 16 according to the Julian calendar. These drawings are classified into two categories. One category features whole-disk drawings for each date, while the other tracks the motion of each specific sunspot group for successive dates.

Consulting this manuscript, we have derived the sunspot group number for each observational date using the Waldmeier classification. We have acquired 103 days of sunspot observations in his drawings and 34 days in his textual descriptions. The latter mostly describes spotless days, whereas three of them report multiple sunspots without exact numerical values. On this basis, we have Hayakawa *et al*.: 2021, Mögling's sunspot observations in 1626 – 1629, *The Astrophysical Journal*, DOI: 10.3847/1538-4357/abdd34identified 134 days with an exact sunspot group number in his manuscript. His background and correspondence show that his observational dates were recorded in the Julian calendar, while the existing datasets misinterpreted this as the Gregorian calendar. Therefore, we have revised the dates of Mögling's sunspot observations. Apart from this calendar issue, we have added 19 days of hitherto overlooked observations, revised three dates (even in the Julian calendar), and excluded eight days of misinterpreted observations. We compared our result with contemporary observations, thereby revising the contemporary observational records of Schickard and Hortensius. Overall, Mögling's revised sunspot group number visualises the declining phase of Solar Cycle −12 and fills the gaps of other contemporary observations. As Mögling probably used a pinhole with a camera obscura, he probably missed all the A-type and B-type sunspot groups, and arguably, the J-type sunspot groups as well. Therefore, his sunspot group number is probably underestimated, overlooking ≥ 33% – 52 % of the total groups (see *e.g*., Carrasco *et al*., 2015).

We have also derived the sunspot positions from Mögling's manuscript. The E-W orientation was set left to right based on the motion of each sunspot group. The N–S orientation was set upside-down based on a comparison of the inclination of the depicted sunspot motion and the calculated B0 angle. On this basis, we derived sunspot positions by combining sunspot-motion drawings for successive days and whole-disk drawings for a specific date. The sunspot groups in 1626 were located 3°–25° in the heliographic latitude in both solar hemispheres. Temporal variations in their latitudinal distributions show their migration to lower heliographic latitudes towards 1629. This result, which agrees fairly well with the existing distributions of contemporary sunspot observations by Scheiner (Arlt *et al*., 2016) and Malapert *et al*. (Carrasco *et al*., 2019b), visualises sunspot migration during the declining phase of Solar Cycle −12. This confirms the hypothesis of the continuity of Solar Cycle −12 from 1621 to somewhere in 1631/1632 (Vokhmyanin and Zolotova, 2018b; Carrasco *et al*., 2019b).

**Acknowledgments**

We thank the Universitäts- und Landesbibliothek Darmstadt for letting us access HS 228 and HS 3020. This work has been supported in part by JSPS Grant-in-Aids JP15H05812, JP20K20918, and JP20H05643, JSPS Overseas Challenge Program for Young Researchers, JSPS Overseas Challenge Program for Young Researchers, the 2020 YLC collaborating research fund, and the research grants for Mission Research on Sustainable Humanosphere from Research Institute for Sustainable Humanosphere (RISH) of Kyoto University and Young Leader Cultivation (YLC) program of12

Nagoya University. BPB acknowledges support by Austrian Science Foundation (FWF) project P 31088 (PI: Ulrich Földsche). We thank Frédéric Clette for helpful discussion on the percentage of Zürich types in the USET. This work has been partly benefitted from discussions in the International Space Science Institute (ISSI, Bern, Switzerland) via the International Team 417 "Recalibration of the Sunspot Number Series", which has been organised by Frédéric Clette and Mathew J. Owens.

**Data Availability**

Mögling's manuscripts are preserved in the manuscript archives of the Universitäts- und Landesbibliothek Darmstadt as HS 228 and HS 3020.